\documentstyle[openbib,fleqn,12pt]{article}
\renewcommand{\baselinestretch}{1.}

\newcommand{\barr}{\begin{array}}
\newcommand{\bea}{\begin{eqnarray}}
\newcommand{\beq}{\begin{equation}}
\newcommand{\ear}{\end{array}}
\newcommand{\eea}{\end{eqnarray}}

\newcommand{\eeq}{\end{equation}}

\begin{document}
\bibliographystyle{unsrt}
{\centering 

{\Large \bf Ergodic properties of quantum conservative systems }
\vskip 2.truecm
{\large   Fausto Borgonovi \\}
{\normalsize\em  Dipartimento di Matematica, Universit\`a Cattolica,\\
via Trieste 17, I--25121 Brescia, ITALY \\ 
Istituto Nazionale di Fisica della Materia , Unit\`a di  Milano\\
Istituto Nazionale di Fisica Nucleare , Sezione di  Pavia \\}

{\large  Giulio Casati  
\\}
{\normalsize\em 
International Center for the study of dynamical systems 
\\
Via Lucini 3, I--22100 Como, ITALY\\
Istituto Nazionale di Fisica della Materia , Unit\`a di  Milano\\
Istituto Nazionale di Fisica Nucleare , Sezione di  Milano \\}
}
\vskip 2.5truecm
{\centerline {\sc \large Abstract}}
{\small
In this paper we discuss the ergodic properties of  quantum 
conservative systems
by analyzing the behavior of two different models.
Despite their intrinsic differencies they 
both show localization effects 
in analogy to the dynamical localization found in 
Kicked Rotator.
}
 
\setcounter{footnote}{0}
\vskip 30pt

\section{Introduction}

The study of quantum systems which are classically chaotic 
is known under the general term of 
Quantum Chaos.
This subject is very interesting and quite new. Indeed, while 
the statistical properties of chaotic classical 
systems are well described by the ergodic theory\cite{Sinai},
much less is known about Quantum Chaos.

Since the preliminary investigations in the 70's   
\cite{CCFI} it became quite clear
that one of the main characteristics of the classical chaotic
motion, namely the diffusion in the phase space, is suppressed by 
quantum mechanics. Despite the 
fact that this phenomenon was discovered in a particular 
 time--periodic Hamiltonian system,
(the so--called Kicked Rotator model)
it was later predicted  and observed in 
other, more physical systems,
like the Hydrogen atom in a  microwave field\cite{CCGS}. 
The 
quantum suppression of   classical 
excitation  has been 
called "dynamical localization"\cite{CIS}.

On the other hand, bounded 
conservative systems, were analyzed from the 
point of view of the statistical properties of eigenvalues 
and eigenvectors\cite{Bohigas}. Important analogies were  
established between the statistical properties of 
Hamiltonian spectra and those derived from  
 random matrices for which  a fairly deep  mathematical analysis 
 had been developed\cite{Mello}. In particular the 
spacings distribution for nearest neighboring levels 
of a classically chaotic system was found to be close
to that derived from  random matrices
belonging to a given 
symmetry  class 
\cite{Bohigas}
 (Gaussian Orthogonal (GOE)
or Gaussian Unitary Ensembles (GUE) 
depending on whether the Hamiltonian 
is invariant or not under 
time reversal). Other approaches
are based on 
 nonlinear $\sigma$-models
and supersymmetric techniques. 

In this paper we investigate two simple conservative systems
in order to show that the dynamical localization can 
appear 
in  this case also.
Roughly speaking, in an ergodic conservative system the motion 
takes place on the whole energy surface.
Correspondingly, in the quantum case, an eigenfunction can 
be extended or localized inside the energy shell.

A main difficulty in understanding quantum chaotic motion
arises from the fact that 
a  bounded conservative quantum system is characterized
by a discrete energy spectrum with a 
 quasi--periodic motion and this can be hardly compared with 
the typical features of the corresponding 
classical chaotic motion
which is characterized by a continuous spectrum.
At first glance this would seem 
a failure of the correspondence principle. 
However the distinction between discrete and continuous 
spectrum becomes meaningful only on infinite time.
On finite time scales, instead, the quantum motion can be 
chaotic as the classical one.

A discussion of  time scales, which play a fundamental r\^{o}le 
in the study of Quantum Chaos, is given  in  Section 2. 
In Section 3 we will analyze a  
model of classically ergodic 
conservative system, the so--called 
Wigner Banded Random Matrix ensemble,
where the dynamical localization phenomenon 
is shown to take place.
Finally, in Section 4,  we will investigate 
the problem of localization in a more physical dynamical system :
the Bunimovich stadium.

\section{Time Scales}

What are the physically meaningful time scales in  quantum 
dynamics?
To answer the question let us 
compare the quantum and classical evolution 
starting  from 
the same initial conditions. Namely, let us 
consider the evolution of a quantum narrow packet 
and of a beam of classical orbits initially centered in the same small
area 
of phase space. 
As it is known, the narrow packet will follow the beam of 
classical orbits as long as the beam remains narrow. 
Due to the exponential instability  of classical chaotic motion, 
this can happen only up to a time $t_E$,
called Ehrenfest time, 
$$
t_E \propto \log (I/\hbar)
$$ 
where $I$ is a 
typical value of the action of the system. 
This time scale, first introduced in \cite{Zasla}, is very short.
Yet, according to the correspondence principle, it grows to infinite
as $\hbar \to 0$.
For times $t<t_E$ the 
 quantum dynamics 
is chaotic as  the classical one,
including the exponential instability of the motion.

Another, more interesting and longer 
time scale, is related to the Heisenberg
uncertainty principle. Indeed, if $\rho^*$ is the level density
of the operative eigenstates involved in the quantum dynamics 
(namely the eigenstates which enter in the initial condition) 
then the discreteness of the spectrum
can be resolved by quantum dynamics only after a
time $t_B$\cite{CIS} :

\beq
t_B \simeq \hbar \rho^*  
\label{breaktime}
\eeq

This  time scale is related to the process of  relaxation
to the quantum steady state.
Indeed, for times $t < t_B$,  
 quantum and classical averages of dynamical observables
will be close to each other while,
 for $t > t_B$ , the quantum system will ``see'' the discrete
nature of the  spectrum 
and consequently it will reach  a stationary regime.
Quite obviously, as required by 
 the correspondence
principle,  this time also goes to infinite when $\hbar \to 0$.
Notice, however, that as $\hbar \to 0$, the time $t_B$ diverges to infinite 
 according to a power law dependence.
Therefore the quantum relaxation time $t_B$ is much larger than the time $
t_E$ which 
characterizes the 
 quantum instability of the motion.
This means that the quantum diffusion and the 
relaxation process take place
in 
absence of exponential instability, which is confined in the small time
interval $t_E$.

The nature of the quantum steady state will depend on the comparison
between  the
 time  $t_B$ and  the ergodic time $t_{erg}$ for the 
 classical relaxation to  equilibrium. Indeed,
if $L$ is a typical length scale over which  classical diffusive 
motion takes place with  diffusion rate $D$, 
then the ergodic time $t_{erg}$ can be estimated as:

\beq
t_{erg} \simeq L^2 / D
\label{ergodic}
\eeq

It is clear that, if the quantum relaxation time $t_B$ 
is larger than the classical one $t_{erg}$, then the quantum 
steady state will be close (apart quantum fluctuations)
to 
 the classical one, 
given by the microcanonical ensemble.
On the other hand, a  more interesting situation appears  when 
$t_B < t_{erg}$. In such  a case, the 
quantum distribution will relax to a steady state which is not
ergodic but localized around the initial excitation.

In conclusion, the nature of the quantum steady state
is determined by the parameter $\lambda$,
which has been called ``ergodicity parameter'' : 

\beq
\lambda^2  = t_B/t_{erg}
\label{lam}
\eeq

Systems characterized by $\lambda \gg 1$ will relax toward a quantum
ergodic steady state, while, for $\lambda \ll 1 $, the quantum stationary
state is localized. 
In the following sections we will discuss, on  specific examples,
the mechanism through which quantum dynamical localization can actually
take place in conservative systems.

\section{The Wigner Band Random Matrices Ensemble}

Apart from  few recent papers\cite{Fyo}, the main interest in the 
Random Matrix Theory 
has been related with  full random matrices
\cite{Mello}. However, for real quantum Hamiltonian systems, matrix 
elements decay
on moving away from the main diagonal and this lead to the consideration
of matrices with a band structure, in which matrix 
elements are different from zero only inside 
a band of size $b$ around the main diagonal.

Let us consider, therefore, the following band random matrix(WBRM) :

\beq
H_{m,n} = \epsilon_n \delta_{m,n} + v_{m,n}
\label{ham}
\eeq

which can be taken to describe an Hamiltonian system of the type $H = H_0 + V$
where $H_0$ is integrable and $V$ is a perturbation which 
renders the Hamiltonian $H$ ergodic and mixing.
The  unperturbed energies 
$\epsilon_n$
are assumed to be distributed 
according to a Poisson law.
The off--diagonal matrix elements
$v_{m,n}$ are taken as Gaussian random numbers with zero average and
variance $v$ 
inside a band of size  $b$:

\beq
\langle v^2_{m,n} \rangle  = v^2  \qquad
{\rm for} \qquad \vert m-n \vert \leq b.
\eeq

Outside the band $b$, the matrix elements are zero. 
This model is then defined by three parameters : $b,v$ and the average
density $\rho$ of unperturbed levels: 

\beq
\rho = {{1}/{\langle \epsilon_n - \epsilon_{n-1}\rangle } }. 
\label{unde}
\eeq

This model, introduced by Wigner\cite{Wigner}, 
has nowadays attracted much  attention 
\cite{Feingold, ccgi}, and several interesting  results have been
obtained.
Here, we only briefly mention those 
connected to 
the localization problem.

 The quantities of interest  are the 
Strength Function or Local Density of States (LDOS):

\beq
w(E \vert E_0 ) = \sum_m \langle \vert \psi_n (E_m) \vert^2 \rangle_n 
\ \delta (E - E_m) 
\label{ldos}
\eeq

and  the averaged eigenfunctions distribution in the energy space:

\beq
W (E_0 \vert E ) = \sum_n \langle \vert \psi_n (E_m) \vert^2 \rangle_m 
\ \delta(E_0 - E_n^0) 
\label{ef}
\eeq

Here, $\psi_n (E_m)$ is the $n$-th component of the eigenfunction having
$E_m$ as eigenvalue and the averages 
($\langle \ldots \rangle_n ,\ \langle \ldots \rangle_m $)
have been done respectively 
over a small number of $n$ values  close to  $E_0$ and over those 
eigenfunctions having an eigenvalue $E_m$ close to $E$.

The non--perturbative regime is defined by the condition that  
the strength of the perturbation must 
be larger than the average unperturbed levels spacing :
$$
\rho v > 1 .
$$

Above the perturbative regime, the shape of LDOS depends on
the Wigner parameter:
$$
q = { {\rho v^2 } \over {b/\rho} }
$$
Here the numerator represents the spreading width induced by the
perturbation
(given by the Fermi golden rule),
 while the denominator stands for the
width, in energy, of the band matrix.

It is possible to show that,  when 
 $q \gg 1$, the LDOS
  has a semicircle form with diameter $\Delta E = 4 v \sqrt{2 b}$
(semicircle   regime (SC) ). In the other regime, characterized by 
  $q \ll 1$, the LDOS has a Lorentzian shape  with the main 
part inside a width $\Gamma  =  2\pi \rho v^2 << \Delta E = 2b/\rho $
(Breit--Wigner   regime (BW) ).

Due to energy conservation, 
the 
number of states occupied by
 an eigenfunction will be bounded from above by
the ergodic localization length $\ell_e \simeq \rho \Delta E$ which 
gives the maximum number of states 
which can be coupled by the perturbation.
 We then consider, as localized, 
 those states 
with a localization length 
 significantly less than the maximum 
 one $\ell_e$, namely we  consider  localization
{\it inside} the energy shell $\Delta E$.
As a measure of the degree of localization we can take, for instance, 
 the inverse participation ratio,  

\beq
\ell = 1/\sum_n \vert \psi_n (E_m) \vert^4
\eeq

It can be shown\cite{ccgi}
that the  ratio $\beta_{loc} = \ell/\ell_e$ obeys the 
following scaling relation: 

\beq
\beta_{loc} = { {\ell}\over {\ell_e} } \approx 1 - e^{-\lambda} 
\label{sca}
\eeq

where the parameter $\lambda$, which 
plays the role here of the ``ergodicity parameter'',
 is defined as 

\beq
\lambda = { {\ell_\infty} \over {\ell_e} } \simeq { {b^2}\over{\rho \Delta E} }
\label{ergod}
\eeq

In Eq. (\ref{ergod}), $\ \ell_\infty$ is the localization 
length for band random matrices with infinite unperturbed 
 density $\rho$\cite{qui,sed}.
The parameter 
$\lambda$ is called ergodicity parameter since, when $\ \lambda \gg 1 \ $,
one has $\ \beta \simeq 1 \ $ and therefore
 $\ \ell \simeq \ell_e \ $. 
On the other side 
when $\ 0 < \lambda \ll 1 \ $ one has 
$ \ \beta \simeq \lambda \ll 1 \ $ and $\ \ell \ll \ell_e \ $.

The global properties of the eigenfunctions are also connected with 
the statistical properties of the spectrum.
Indeed, it was found that effects of localization 
manifest in 
the repulsion of neighboring levels. To be more precise, one can 
show\cite{felixrep} that, in case of localization,  
the  distribution of neighboring levels spacing 
obeys the following relation:
\beq
p(s) = A s^\beta \, \exp \, \big[ -\pi^2 \beta s^2/16 - 
(B- \pi \beta/4) \, s \, \big] 
\label{felo}
\eeq
where $A,B$ are constants obtained from normalization. 
An interesting fact, which so far has not yet received a
theoretical explanation,
is that the numerical value of the repulsion parameter $\beta$
 in (12) turns out to be very close 
to the localization parameter 
$\beta_{loc}$\cite{ccgi}.

\section{The Bunimovich Stadium}

In the previous section we have shown that quantum dynamical localization 
can take place in the model of WBRM. Even though
WBRM are believed to describe the qualitative 
properties of conservative, classically chaotic, Hamiltonian 
systems, it is highly desirable to analyze a
more realistic Hamiltonian system. To this end,
two-dimensional billiards are very convenient objects to study
since they have very clean mathematical properties,
from complete integrability (e.g. the circle)
to complete chaotic motion (e.g a dispersive billiard).
Moreover their classical and quantum dynamics can be numerically studied 
with sufficient good accuracy. Finally, modern laboratory techniques
allow for quite accurate experimental investigations. 

In the following we consider a well known billiard model, the so--called 
 Bunimovich Stadium, which consists
of two semicircles, with radius $R=1$, connected by two straight lines
with length $2a$. Inside this bounded two--dimensional region we
consider the motion 
of a point particle with mass $m=1$ colliding elastically with the
boundary. The classical dynamics depends only on the ratio
$\epsilon = a/R$ and it can be rigorously proven to be ergodic 
and mixing for
any $\epsilon \ne 0 $. 
When $\epsilon \sim 1 $ the relaxation time to statistical  equilibrium 
is very short, just few collisions with the
boundary. Here we are  interested to the case $\epsilon \ll 1 $
when the stadium is very close to the circle.
For  the billiard in a circle, 
there exist two 
constants of motion,  the energy $E= m{\vec{v}}^2/2$
and the angular momentum $\vec{l} = l_z \hat{k}$
(here $\hat{k}$ is the unit vector perpendicular to the plane of the 
stadium and identifying the $z$ axis).
For $\epsilon > 0$ the rotational
symmetry around the $z$-axis  
is broken and $l_z$ is not a
constant of motion any more. 
Nevertheless, if $\epsilon$ is sufficiently
small, one can expect that the angular momentum will change slowly
in time.
Indeed, the angular momentum can vary in the interval 
$ \vert l_z \vert < 
l_{max}=m (R+a ) \vert \vec{v} \vert  \simeq \sqrt{2 m R^2 E}$;
now, 
if for example at $t=0$ the angular momentum is zero, then 
it will evolve in time in a diffusive way, with diffusion coefficient D, 
until the system will reach the 
equilibrium state 
given by the microcanonical ensemble. The ergodic time $t_{erg}$,
namely the time necessary to reach the statistical equilibrium, can be
estimated as:

\beq
t_{erg} \simeq   l_{max}^2 / D   
\eeq

The diffusion coefficient D can be computed numerically
\cite{BCL}.
In terms 
of 
the rescaled angular momentum
$L= l_z/l_{max}$, 
the diffusion coefficient, in the number of collisions
with the boundary, is given by 
$D_0 = 1.5 \  \epsilon^{5/2}$. 
By taking into account that the time interval between two 
successive collisions is $t_c \sim E^{-1/2}$ 
one has that, 
neglecting numerical constants ($m=R=1$) , 

\beq
t_{erg}  \sim { {E} \over { E \epsilon^{5/2}}} t_c 
\sim \epsilon^{-5/2} E^{-1/2}
\label{et} 
\eeq

The classical dynamics of the billiard can be approximated 
by the following area--preserving map which gives the change
of $L$ and of its related conjugated variable $\theta$, between
two successive collisions with the boundary\cite{BCL}:

\begin{equation}
\begin{array}{l}
\bar{L} = L + \varepsilon \sin \theta \  {\rm sgn} (\cos \theta )
\nonumber\\
\bar{\theta} = \theta +\pi -2 \arcsin{\bar{L}} \\
\label{mapc}
\end{array}
\end{equation}

Here  $\varepsilon =-2\, \epsilon \, {\rm sgn} (L_0) \sqrt{1-L_0^2}$ and
$L_0$ is the initial value of the angular momentum.

The map (\ref{mapc})
represents a first order approximation to the real dynamics
when the initial rescaled angular momentum $L_0$
is not too large.
It also give rise, for $\varepsilon << 1$, 
to a diffusive motion with a diffusion rate $D \propto \varepsilon^{5/2}$
in agreement with numerical computations on the real billiard. 

The map description 
allows to understand 
the properties of 
the quantum dynamics 
in analogy to 
 the Kicked 
Rotator model\cite{CCFI}. Indeed we know that, 
above the quantum 
perturbative regime $\varepsilon > \hbar$, quantum dynamics   
will follow the classical 
diffusive motion up to a certain time, called break--time,
(see also Eq.(\ref{breaktime})) given by  $\tau_b \sim D_0$\cite{CIS},
where $\tau_b$ is measured in the number of collisions and $D_0$ is
the dimensionless diffusion coefficient. In physical units, this time 
  is given by:

\beq
t_B \sim  \epsilon^{5/2} E^{1/2}/\hbar^2
\eeq

This allows to estimate the ergodicity parameter
$\lambda$  (Eq.(\ref{lam}))

\beq
\lambda^2 = t_B/t_{erg} =  E \epsilon^5/\hbar^2
\label{ergo2}
\eeq

The  value $\lambda = 1$ 
gives the critical energy value:

\beq
E_{erg} \sim \hbar^2/\epsilon^5
\eeq

above which we expect quantum ergodic behavior.
On the contrary, for energies $E < E_{erg}$,
 we expect dynamical localization.
Since the 
total average 
 number of  states up to the energy $E$\cite{Bohigas} is given by 

\beq
\langle {\cal N}(E)\rangle \approx { {m {\cal A} }\over {2\pi\hbar^2 }} E
\label{weyl}
\eeq

where ${\cal A}$ is the area of the billiard, 
 one can estimate the level number 
 $N_{erg} \sim \epsilon^{-5}$
above which quantum  ergodic behavior is expected.
Numerical computations
show that, in agreement with the above estimates,
 only above $N_{erg}$ the  levels spacing distribution 
 is very close to a Wigner--Dyson distribution,
characteristic of the 
GOE ensemble.

One can also numerically compute 
 the repulsion parameter
$\beta$, using Eq.(\ref{felo})
and 
one finds \cite{futuro}
\beq
\beta \simeq 1 - e^{-\lambda}
\eeq
where $\lambda$ is the ergodicity parameter
(\ref{ergo2}).
This quite surprising  result
reinforces the similarity between WBRM and the quantum behavior
of classically chaotic systems.

Finally , 
 we would like to notice that the ergodicity
parameter $\lambda$ 
turns out be proportional to 
the dimensionless conductance:

\beq
g = { { E_c} \over {\Delta} }
\eeq

where $E_c$ is the Thouless energy and $\Delta$ is the average levels
spacing. Indeed taking into account that 

\beq
E_c \simeq { {\hbar}\over{t_{erg}}}
\eeq

and

\beq
\Delta \simeq { {\hbar}\over{t_H} }
\eeq

where $t_H \simeq \hbar {{{\cal N}(E)}\over{E}} \simeq \hbar^{-1} $   
(see Eq.(\ref{weyl}))
is the Heisenberg time; one gets 
\beq
g = { {t_H}    \over {t_{erg}                  } } \simeq 
    { {1/\hbar}\over {\epsilon^{-5/2} E^{-1/2} } } \propto \lambda.
\eeq

\section{Conclusions}

In this paper we have discussed the problem of localization in
conservative, classically chaotic, Hamiltonian systems. The existence
of localization in such systems would restrict quantum distributions
to smaller regions of phase space than classically allowed, and would
therefore introduce significant deviations from ergodicity.
As surmised in \cite{cas}, this lack of quantum ergodicity  may lead to
interesting consequences for quantum equilibrium statistical
distributions.

\renewcommand{\baselinestretch} {1}

\vfill\eject

\end{document}